\documentclass[12pt]{iopart}

\usepackage{graphicx}
\def\PT{p_\mathrm{T}}
\def\auau{Au~+~Au }
\def\pp{$p$~+~$p$ }

\def\snn{\sqrt{s_\mathrm{NN}}}
\def\raa{R_\mathrm{AuAu}}
\def\taa{\langle T_\mathrm{AuAu} \rangle}
\def\npart{N_\mathrm{part}}

\begin{document}
%%%%%%%%%%%%%%%%%%%%%%%%%%%%%%%%%%%%%%%%%%%%%%%%%%%%%%%%%%%%%%%%%%%%%%%%%%%%%%%%%%%%%%%%%%%%%%%%%%%%
\title[Heavy Quark Measurements by Single Electrons in the PHENIX Experiment]
	  {Heavy Quark Measurements by Single Electrons\\ in the PHENIX Experiment}

\author{F. Kajihara (for the PHENIX Collaboration)\footnote[7]{For the full PHENIX Collaboration 
		author list and acknowledgments, see appendix `Collaborations' of this volume.}}

\address{Center for Nuclear Study (CNS), 
         Graduate School of Science, the University of Tokyo\\
		 CNS in RIKEN, 2-1 Hirosawa, Wako, Saitama, 351-0198, Japan}
\ead{kajihara@bnl.gov}
%%%%%%%%%%%%%%%%%%%%%%%%%%%%%%%%%%%%%%%%%%%%%%%%%%%%%%%%%%%%%%%%%%%%%%%%%%%%%%%%%%%%%%%%%%%%%%%%%%%%

%%%%%%%%%%%%%%%%%%%%%%%%%%%%%%%%%%%%%%%%%%%%%%%%%%%%%%%%%%%%%%%%%%%%%%%%%%%%%%%%%%%%%%%%%%%%%%%%%%%%
\begin{abstract}
Transverse momentum ($\PT$) distribution of electrons for $0.3<\PT<9.0$ GeV/$c$ have been measured in
midrapidity ($|\eta|<0.35$) in \auau collisions and \pp collisions at $\snn = 200$ GeV by the 
RHIC-PHENIX experiment. Two methods for background subtraction were applied to determine the electron
yield from open charm and bottom decays. The nuclear modification factor was calculated, and 
significant suppression at high-$\PT$ was observed in \auau collisions, indicating substantial energy
loss of heavy quarks in the dense medium. 
\end{abstract}
%%%%%%%%%%%%%%%%%%%%%%%%%%%%%%%%%%%%%%%%%%%%%%%%%%%%%%%%%%%%%%%%%%%%%%%%%%%%%%%%%%%%%%%%%%%%%%%%%%%%

%%%%%%%%%%%%%%%%%%%%%%%%%%%%%%%%%%%%%%%%%%%%%%%%%%%%%%%%%%%%%%%%%%%%%%%%%%%%%%%%%%%%%%%%%%%%%%%%%%%%
\section{Introduction}
Heavy quark (charm and bottom) measurement is important to extend our knowledge of underlying QCD 
properties. Its measurement in \pp collisions at RHIC serves as a good test of QCD. Bottom productions
at the Tevatron and HERA are well described by next-to-leading order (NLO) perturbative QCD 
calculations~\cite{bib:00}. RHIC data will also help our theoretical understanding of the production 
and fragmentation mechanism of heavy quarks, especially, charm.  In \auau collisions at RHIC, strong
suppressions  of  light flavor mesons  at high transverse momentum ($\PT$) have been observed
\cite{bib:01,bib:02,bib:03}. The suppressions are due to significant energy loss of partons through 
an extremely dense matter, not conventional hadronic matter. Due to the large mass, heavy quark can
interact with the dense medium in different ways from light partons. The measurement will provide
significant and complementary information of mechanism of parton energy loss. 
%%%%%%%%%%%%%%%%%%%%%%%%%%%%%%%%%%%%%%%%%%%%%%%%%%%%%%%%%%%%%%%%%%%%%%%%%%%%%%%%%%%%%%%%%%%%%%%%%%%%

%%%%%%%%%%%%%%%%%%%%%%%%%%%%%%%%%%%%%%%%%%%%%%%%%%%%%%%%%%%%%%%%%%%%%%%%%%%%%%%%%%%%%%%%%%%%%%%%%%%%
\section{Experiment and Analysis}
The PHENIX experiment took high statistic data at $\snn=200$ GeV in Run-4 \auau collisions (2004) 
\cite{bib:04} and in Run-5 \pp collisions (2005) \cite{bib:05}. Electrons ($e^+/e^-$) in $0.3<\PT<9.0$
GeV/$c$ were detected by two PHENIX central arms, each covering azimuthal angle $\Delta \phi = \pi/2$
and pseudo-rapidity $|\Delta \eta|<0.35$~\cite{bib:06}. 

All measured electrons can be categorized into two groups. The first group consists of 
``\textit{photonic}'' electrons which mainly come from Dalitz decays of mesons ($\pi^0, \eta$, etc.)
and photon conversion. The second group is termed as ``\textit{non-photonic}'' electrons in which
open charm/bottom decays are dominant sources. We applied two methods, ``\textit{cocktail}'' and
``\textit{converter}'' methods to extract non-photonic electrons by subtracting photonic electrons
from all electrons~\cite{bib:04,bib:05}. The both methods give the consistent result. 
Cocktail method is effective to count signal electrons at high-$\PT$, where the signal to background
ratio is large. 
Converter method has small systematic error by the direct measurement of photonic electron components, 
but the statistical error is dominated by the statistics of converter-installed runs. Almost all the
non-photonic electrons are produced from semileptonic decays of open charms and bottoms. Backgrounds
in the non-photonic electrons come from mainly weak decays of Kaon ($K_{e3}$) and di-electron decays
(vector meson decays and Drell-Yan process). Kaon contributes $<10~\%$ at $\PT = 0.5$ GeV/$c$ compared
to photonic electron yields, while vector mesons are very small, and Drell-Yan process is negligible 
in our measurable $\PT$ range. Since Kaon and vector mesons have been already measured by the PHENIX,
backgrounds from those sources can be evaluated with simulations and subtracted from non-photonic
electrons. 
%%%%%%%%%%%%%%%%%%%%%%%%%%%%%%%%%%%%%%%%%%%%%%%%%%%%%%%%%%%%%%%%%%%%%%%%%%%%%%%%%%%%%%%%%%%%%%%%%%%%

%%%%%%%%%%%%%%%%%%%%%%%%%%%%%%%%%%%%%%%%%%%%%%%%%%%%%%%%%%%%%%%%%%%%%%%%%%%%%%%%%%%%%%%%%%%%%%%%%%%%
\begin{figure}[htbp]
\begin{minipage}{0.5\textwidth}
\begin{center}
\includegraphics[width=\textwidth]{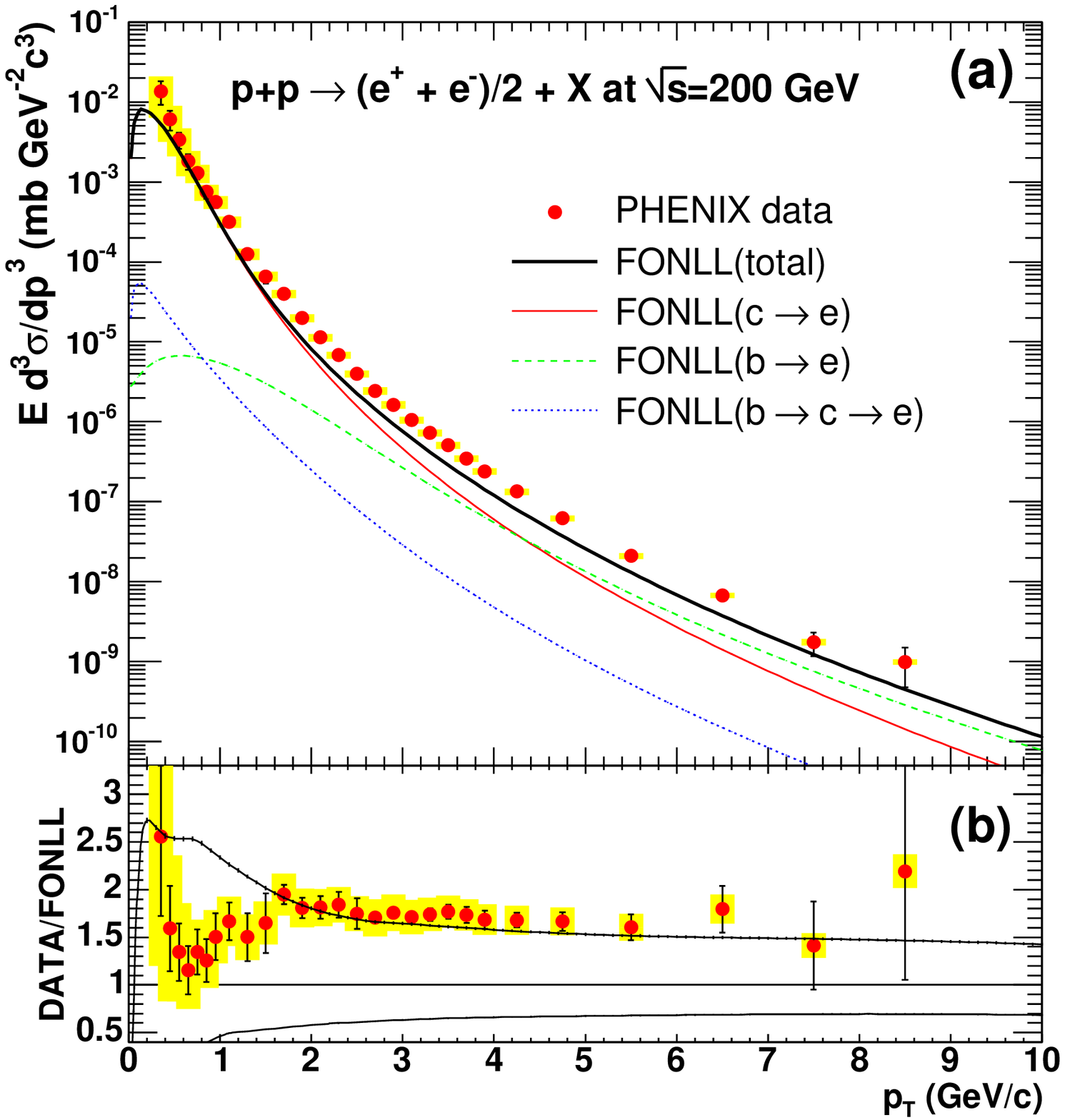}
\end{center}
\end{minipage}
\hspace{2mm}
\begin{minipage}{0.45\textwidth}
\begin{center}
\includegraphics[width=\textwidth]{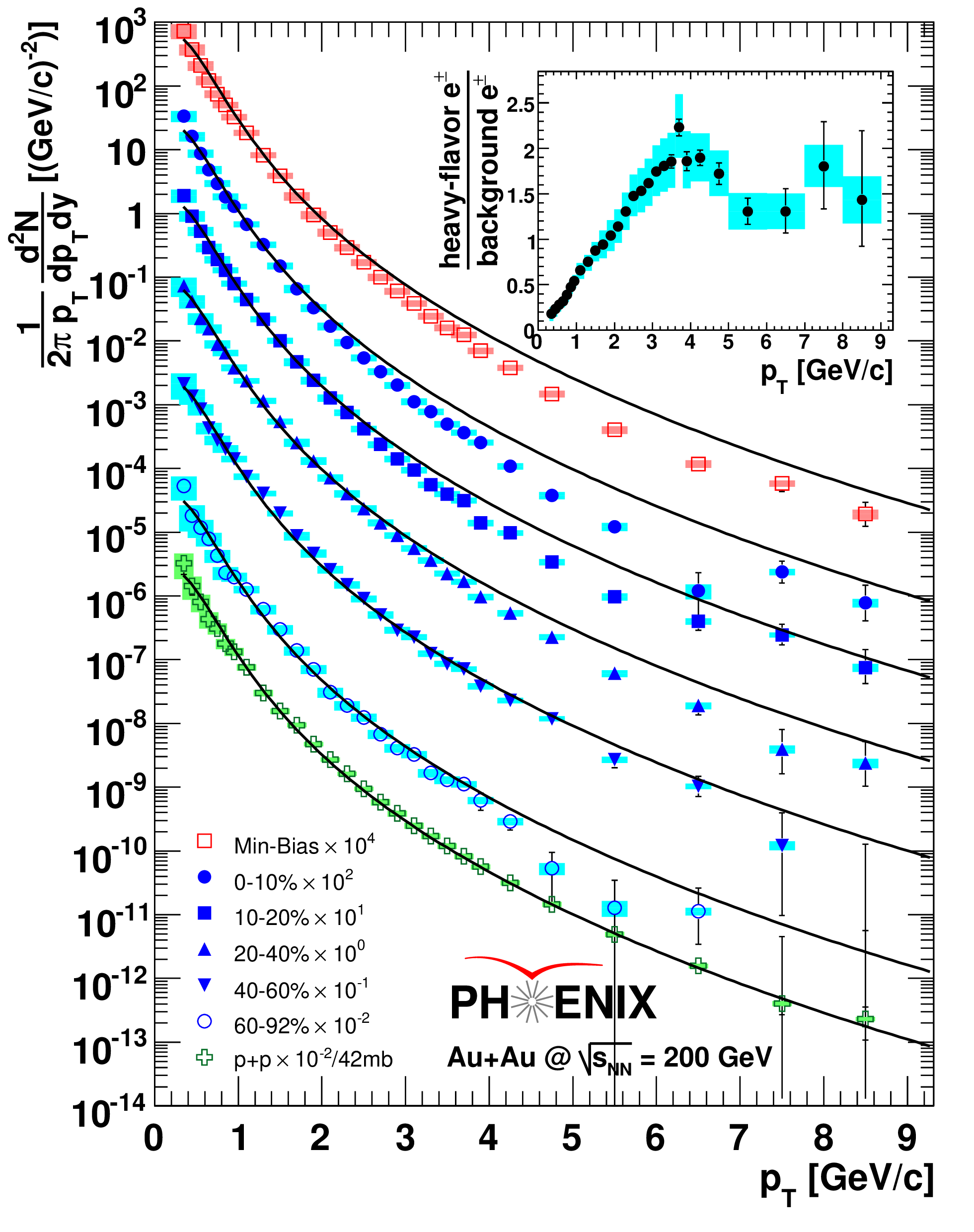}
\end{center}
\end{minipage}

\hspace{-2cm}
\begin{minipage}{0.45\paperwidth}
\caption{(a) Invariant differential cross sections of electrons from semileptonic decays of heavy 
		 quarks in Run-5 \pp collisions. The error bars (bands) represent the statistical (systematic)
		 errors. The curve are the FONLL calculations. (b) Ratio of the data and the FONLL calculation.
		 The upper (lower) curve shows the theoretical upper (lower) limit of FONLL calculation. 
		 \label{fig:01}}
\end{minipage}
\hspace{-2cm}
\begin{minipage}{0.45\paperwidth}
\caption{Invariant differential yields of electrons from semileptonic decays of heavy quarks in Run-5 
		 \pp and Run-4 \auau collisions for each centrality-class, scaled by power of ten for clarity. 
		 The solid lines are the result of the FONLL calculation normalized ($\times 1.71$) to the \pp 
		 data and scaled with $\taa$ for each \auau data. The inserted box shows the signal/background
		 ratio of electrons for minimum bias events in \auau collisions. Error bars (boxes) depict
		 statistical (systematic) uncertainties.\label{fig:02}}
\end{minipage}
\end{figure}
%%%%%%%%%%%%%%%%%%%%%%%%%%%%%%%%%%%%%%%%%%%%%%%%%%%%%%%%%%%%%%%%%%%%%%%%%%%%%%%%%%%%%%%%%%%%%%%%%%%%

%%%%%%%%%%%%%%%%%%%%%%%%%%%%%%%%%%%%%%%%%%%%%%%%%%%%%%%%%%%%%%%%%%%%%%%%%%%%%%%%%%%%%%%%%%%%%%%%%%%%
\section{Results}
Figure~\ref{fig:01} (a) shows the invariant differential cross section of electrons from heavy flavor
decays~\cite{bib:04}. The data are compared with a fixed-order-plus-next-to-leading-log (FONLL) pQCD
calculation~\cite{bib:07}. In Fig.~\ref{fig:01} (b), the ratio of data to the FONLL is shown ($1.71 
\pm 0.02^\mathrm{~stat} \pm 0.18^\mathrm{~sys}$). The upper limit of the FONLL calculation is compatible
with the data. Total charm cross section is derived by integrating the heavy-flavor electron cross
section for $\PT>0.4$ GeV/$c$: $d\sigma/dy = 5.95\pm 0.59^\mathrm{~stat} \pm 2.0^\mathrm{~sys}~\mu$b.
    
\begin{figure}[htbp]
\hspace{-3cm}
\begin{minipage}{\paperwidth}
\begin{center}
\includegraphics[width=0.3\paperwidth]{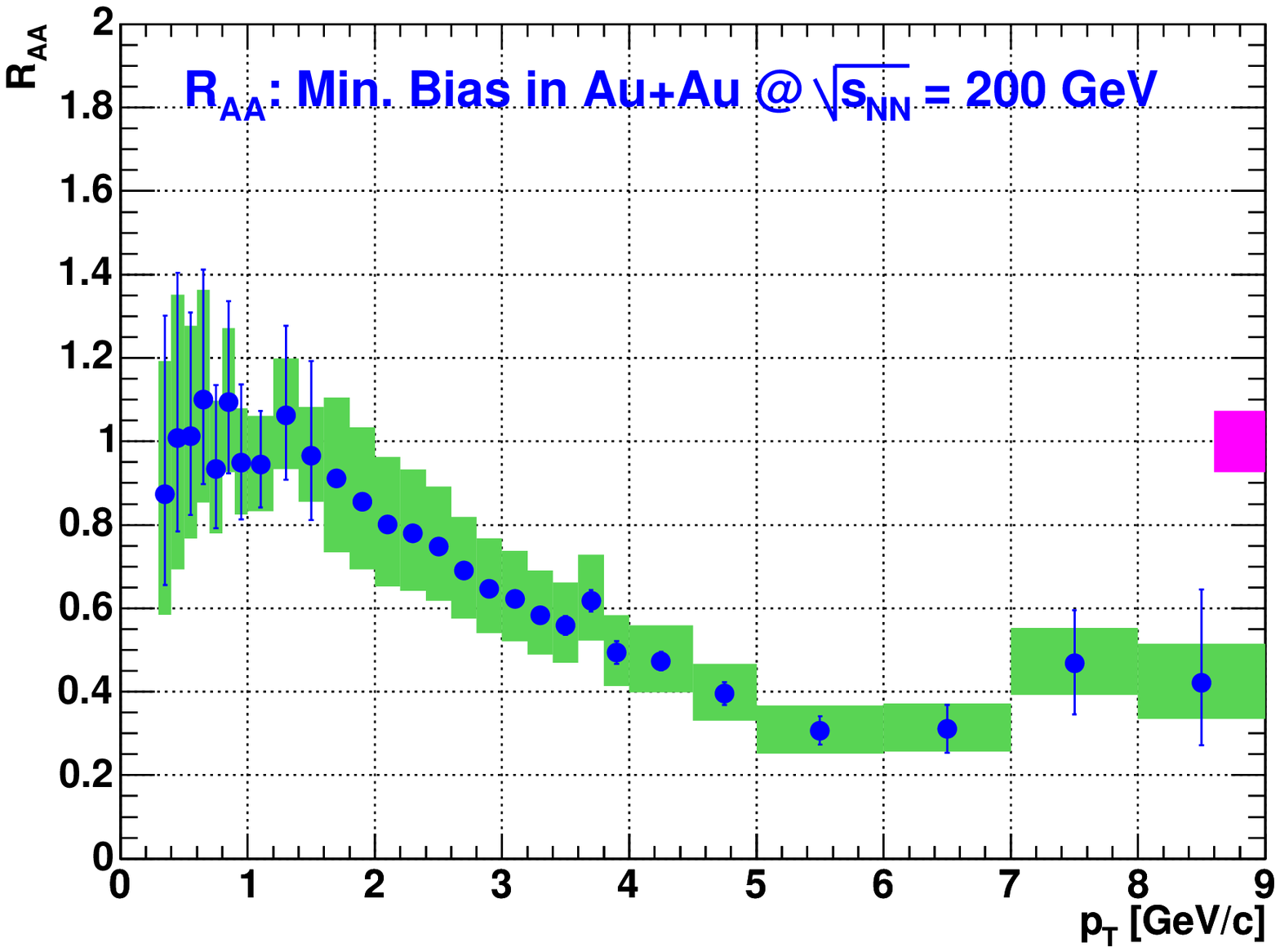}
\includegraphics[width=0.3\paperwidth]{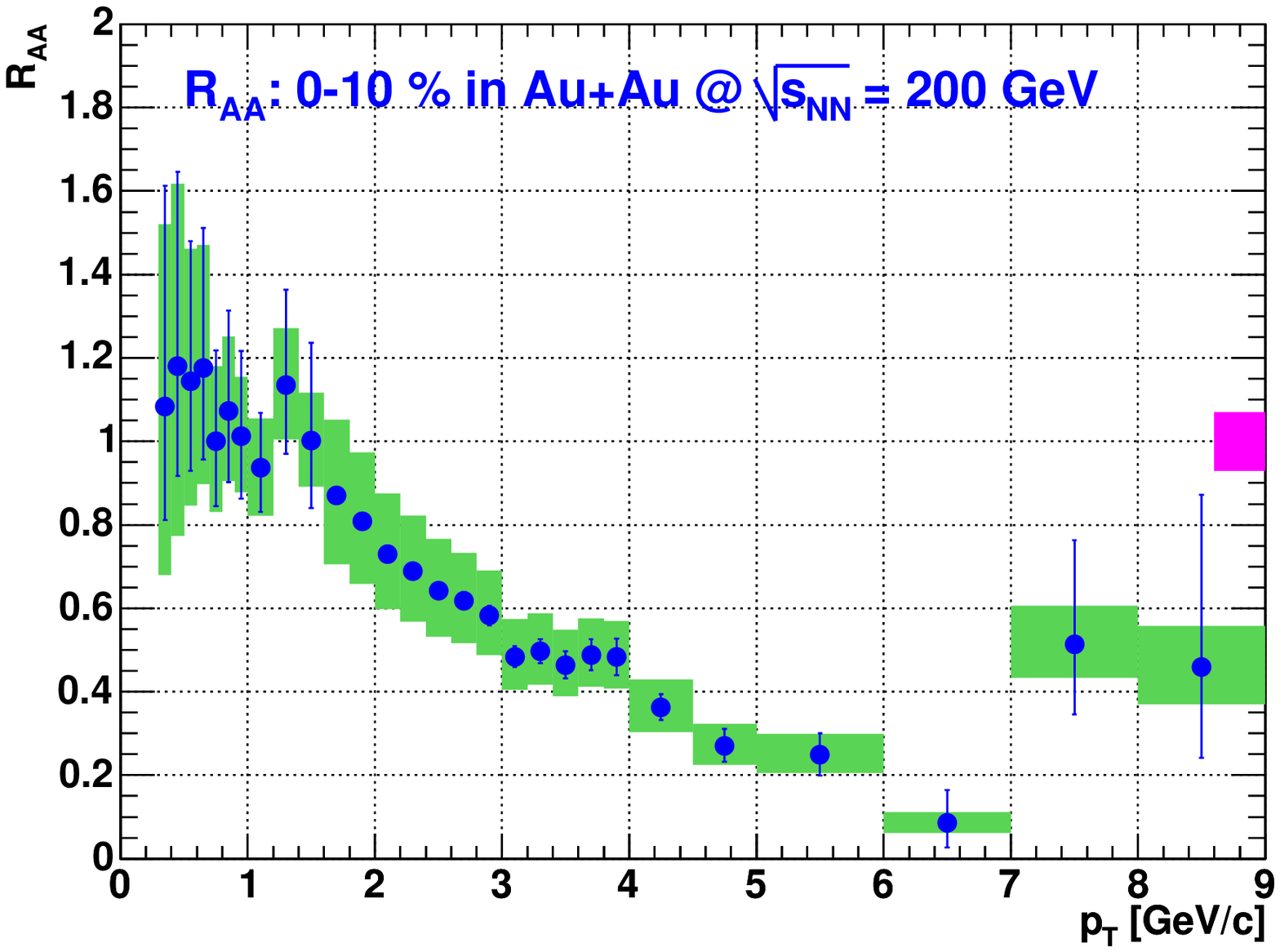}
\includegraphics[width=0.3\paperwidth]{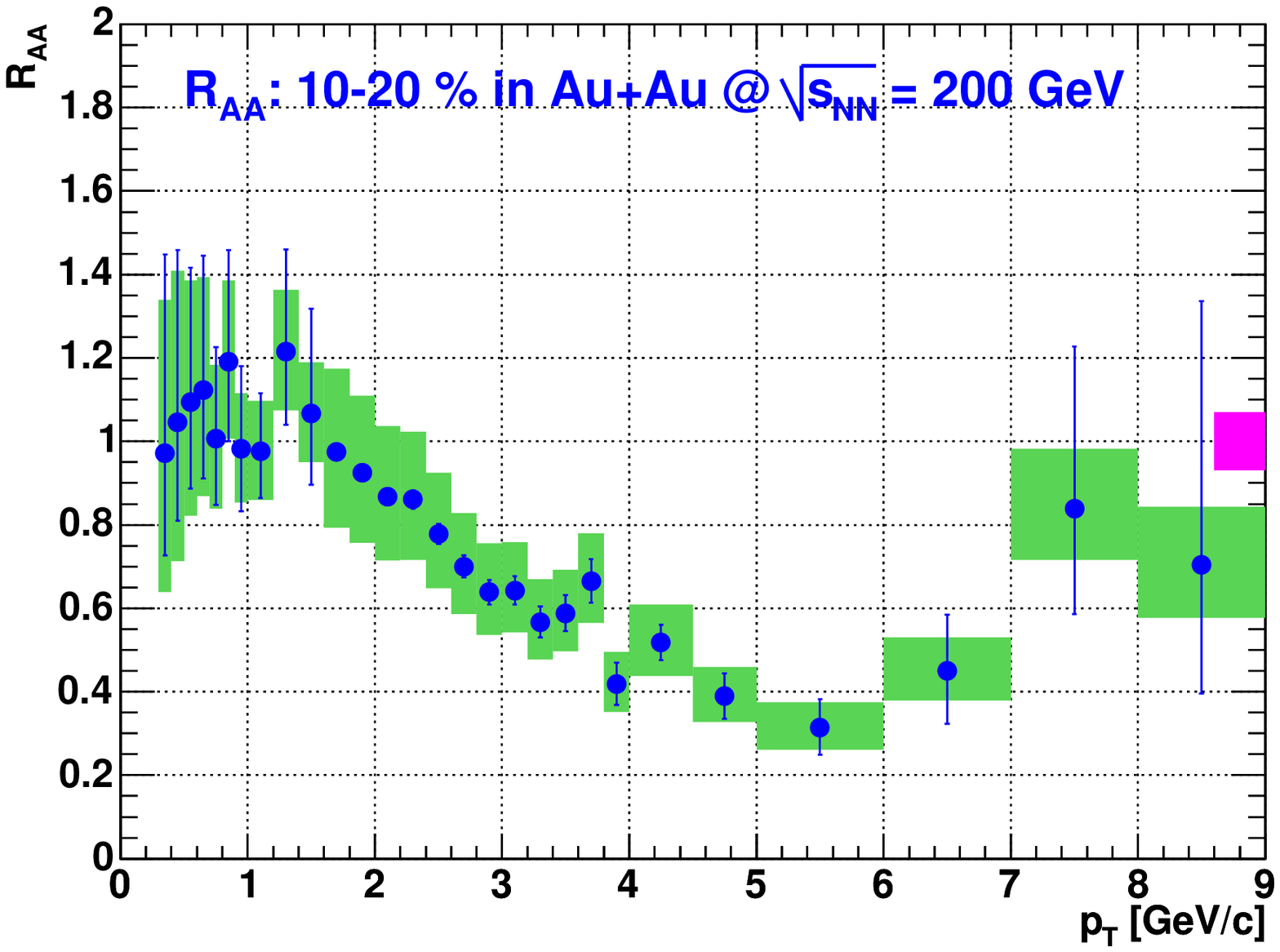}
\end{center}
\end{minipage}

\hspace{-3cm}
\begin{minipage}{\paperwidth}
\begin{center}
\includegraphics[width=0.3\paperwidth]{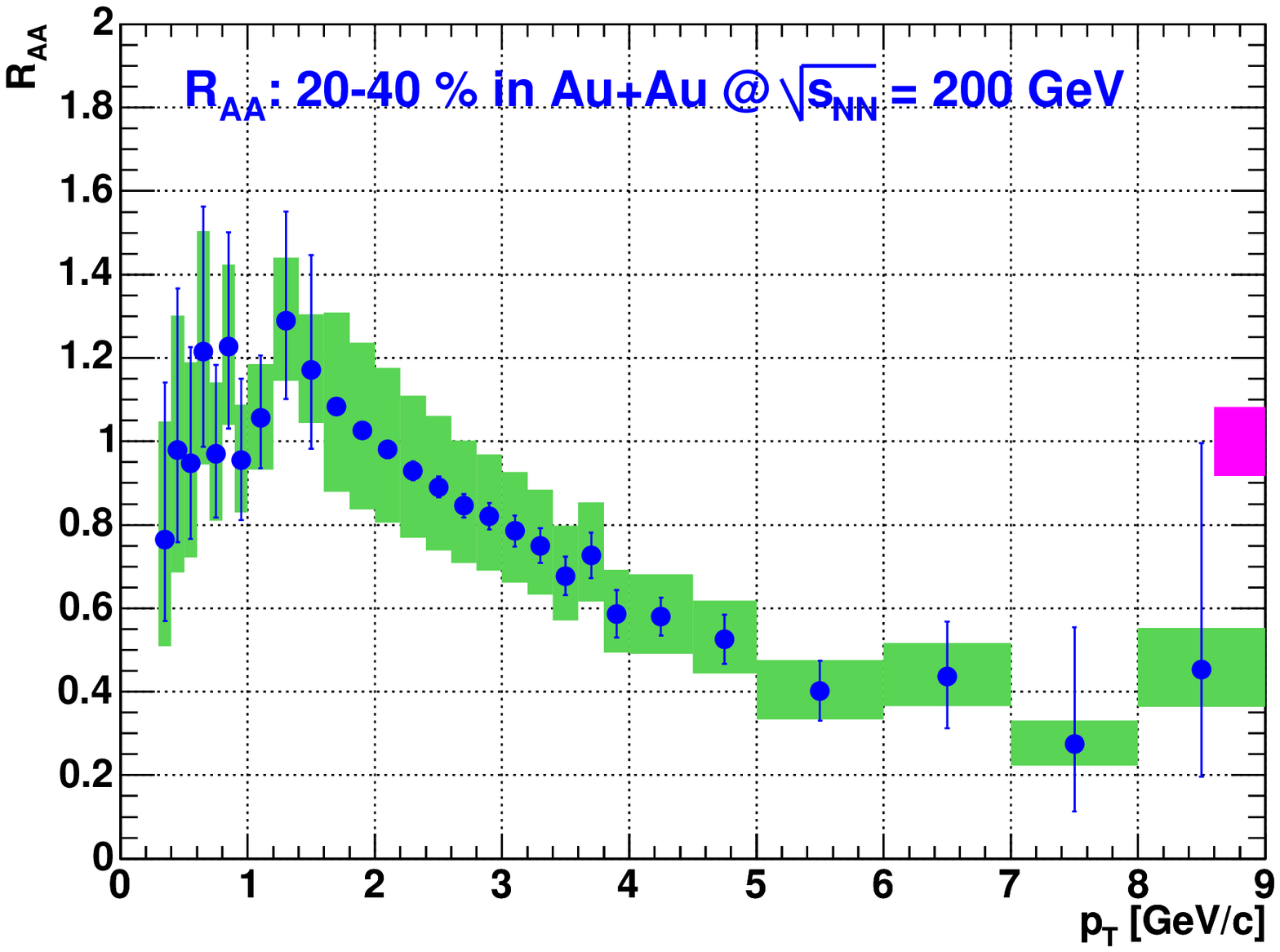}
\includegraphics[width=0.3\paperwidth]{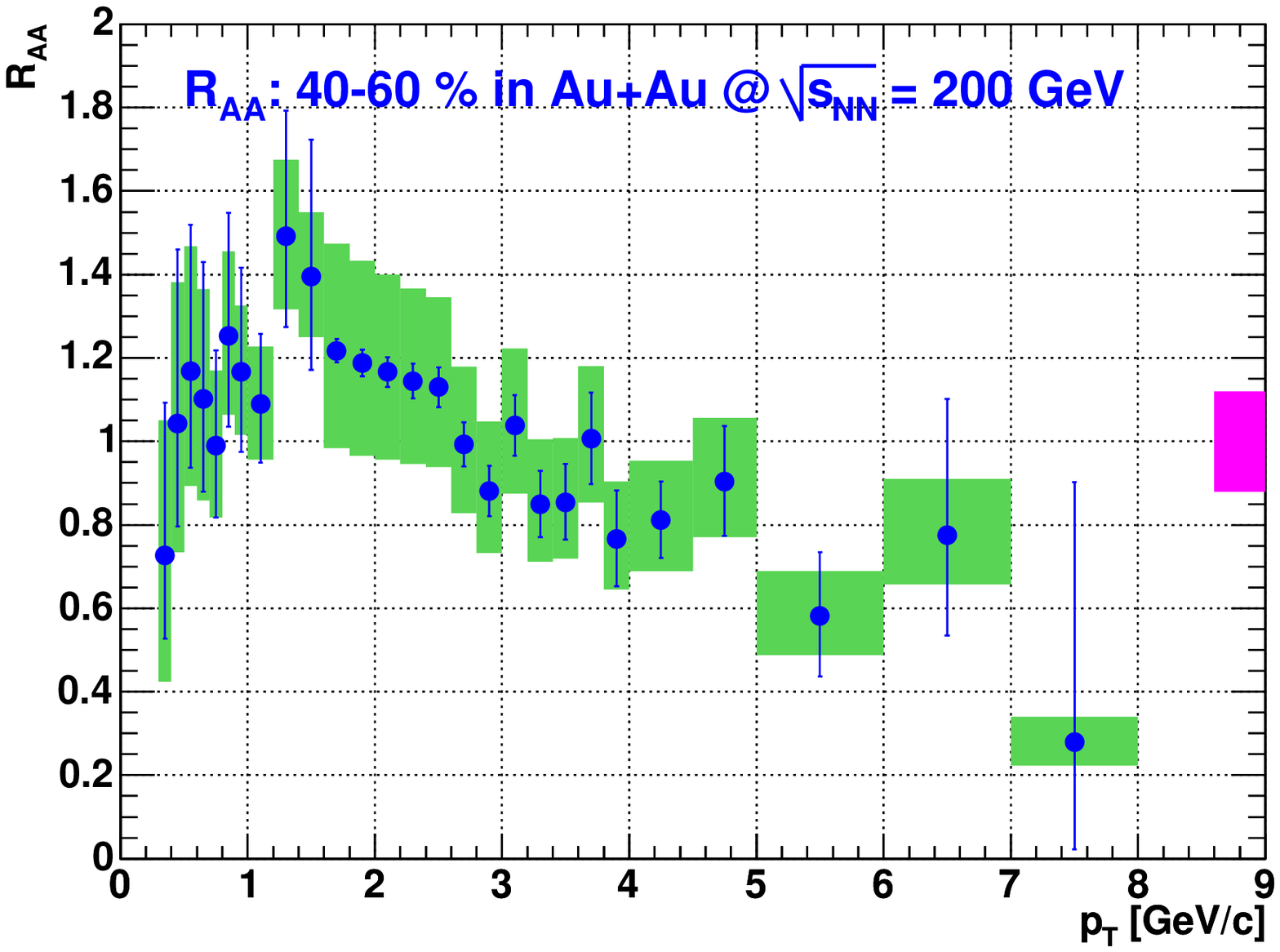}
\includegraphics[width=0.3\paperwidth]{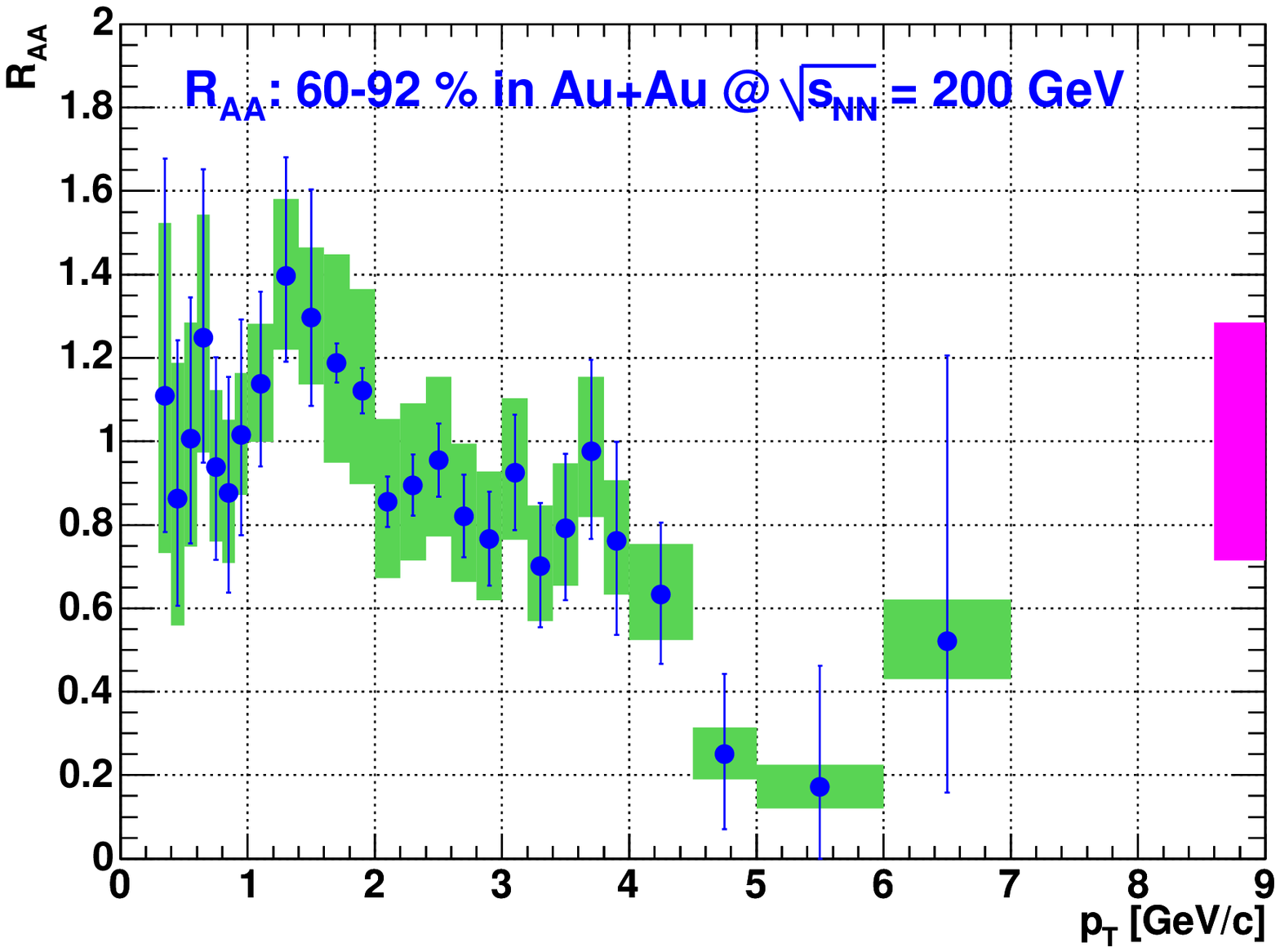}
\end{center}
\end{minipage}
\caption{$\raa$ of heavy-flavor electrons as a function of $\PT$ in each centrality class (minimum
		 bias, 0-10 \%, 10-20 \%, 20-40 \%, 40-60 \%, and 60-92 \%). The error bars (boxes) dipict 
		 point-by-point (scaling) uncertainties of combined statistical and systematic errors,
		 derived from \auau and \pp data. The box around $\raa=1$ in the right side shows the 
		 uncertainty of $\taa$. \label{fig:03}}

\vspace{0.5cm}
\begin{minipage}{0.5\textwidth}
\begin{center}
\centering{\includegraphics[width=\textwidth]{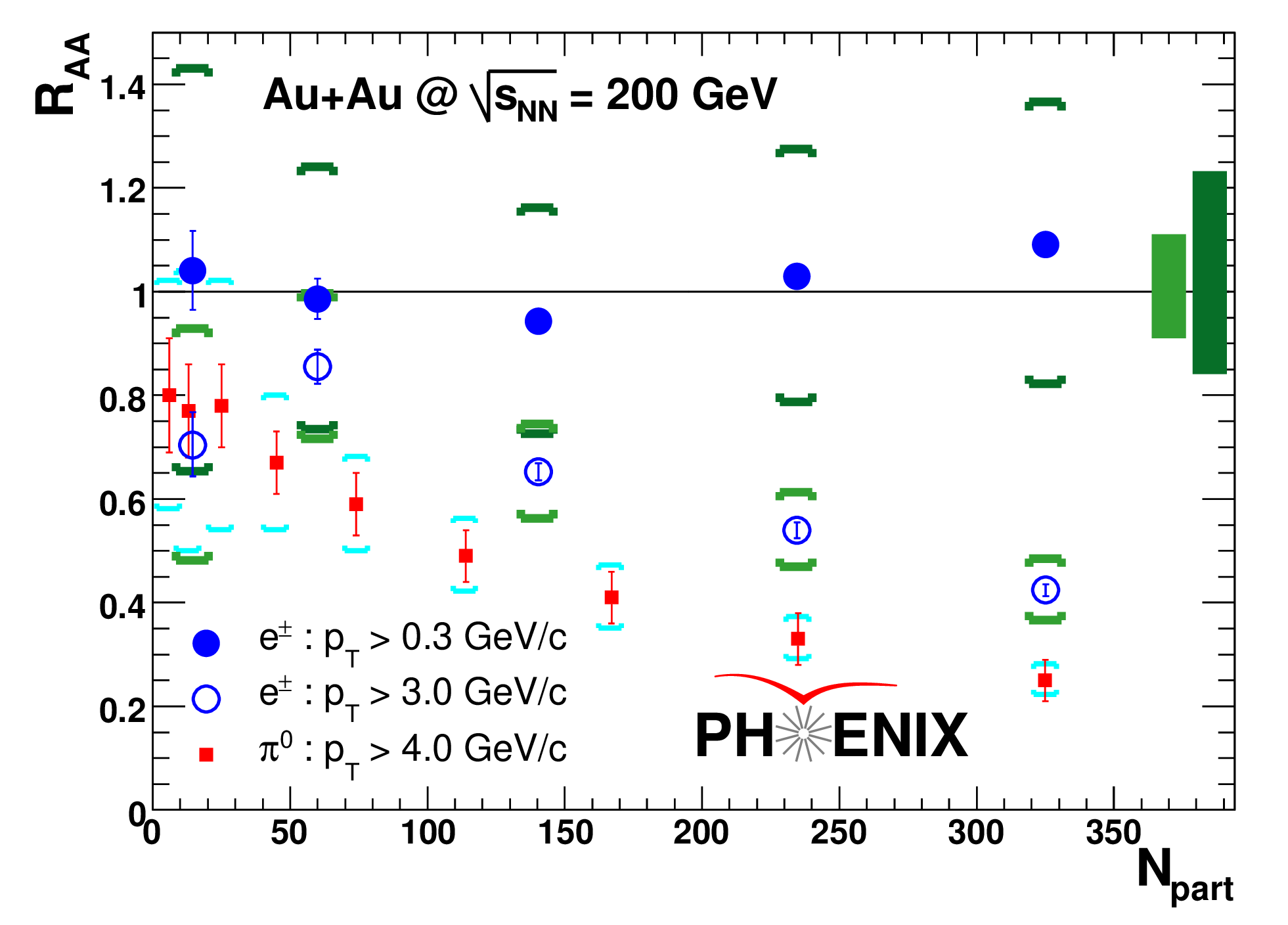}}
\end{center}
\end{minipage}
\hspace{-1.5cm}
\begin{minipage}{0.6\textwidth}
\begin{center}
\caption{$\raa$ of heavy-flavor electrons with $\PT$ above 0.3 and 3.0 GeV/$c$. and of $\pi^0$
		 with $\PT > 4$ GeV/$c$ as a function of $\npart$. Error bars (brackets) depict statistical
		 (point-by-point systematic) uncertainties. The right (left) box error around $\raa=1$ in the 
		 right side shows the relative uncertainty from the \pp data common to the all points except
		 for $\pi^0$ data. 
		 \label{fig:04}}
\end{center}
\end{minipage}
\end{figure}

Figure~\ref{fig:02} shows invariant differential yields of electrons from heavy flavor decays for each 
centrality-class in \auau collisions~\cite{bib:04}. To quantify the suppression, the nuclear 
modification factor, $\raa(\PT)$ was calculated:\\
%%%%%%%%%%%%%%%%%%%%%%%%%%%%%%%%%%%%%%%%%%%%%%%%%%%%%%%%%%%%%%%%%%%%%%%%%%%%%%%%%%%%%%%%%%%%%%%%%%%%
\begin{eqnarray}
  \raa(\PT) = \frac{dN^\mathrm{AuAu}/d\PT}
									   {\taa d\sigma^{pp}_\mathrm{in}/d\PT}
								= \frac{dN^\mathrm{AuAu}/\PT}
									   {N_\mathrm{col}dN^{pp}/d\PT}.
\end{eqnarray}
%%%%%%%%%%%%%%%%%%%%%%%%%%%%%%%%%%%%%%%%%%%%%%%%%%%%%%%%%%%%%%%%%%%%%%%%%%%%%%%%%%%%%%%%%%%%%%%%%%%%
Here, $\sigma^{pp}_\mathrm{in}$ is the inelastic scattering cross section of \pp, $\taa$ is the 
nuclear thickness function for \auau and $N_\mathrm{col} = \taa \cdot \sigma^{pp}_\mathrm{in}$.
$\raa(\PT)$ for each centrality-class is shown in Fig.~\ref{fig:03}. 
Very strong suppression is clearly seen at high-$\PT$ from the most central to the mid
central collisions, comparable to the suppression observed for $\pi^0$ and $\eta$~\cite{bib:01,bib:03}. 
Figure~\ref{fig:04} shows $\raa$ as a function of the number of participant ($\npart$). For $\PT>0,3$
GeV/$c$, $\raa$ is close to unity for all $\npart$, which indicates binary scaling of the total 
heavy-flavor yield works.  For $\PT>3$ GeV/$c$, the $\raa$ decreases systematically with $\npart$.

%%%%%%%%%%%%%%%%%%%%%%%%%%%%%%%%%%%%%%%%%%%%%%%%%%%%%%%%%%%%%%%%%%%%%%%%%%%%%%%%%%%%%%%%%%%%%%%%%%%%
\section{Summary and Outlook}
The PHENIX has measured electrons from semileptonic decays of heavy quarks in RHIC Run-4 \auau and
Run-5 \pp collisions at $\sqrt{s_{NN}}=200$ GeV. The FONLL calculation agrees with \pp data within
the theoretical and experimental uncertainties. The nuclear modification factor ($\raa$) shows a very
strong suppressive effect. The result suggests that even heavy quarks lose their energy in high dense
medium. To understand it systematically, we need to separate the contributions from charm or bottom
in the future experiment. The mass difference will provide us the more information of the energy-loss
mechanism in the high dense matter.
%%%%%%%%%%%%%%%%%%%%%%%%%%%%%%%%%%%%%%%%%%%%%%%%%%%%%%%%%%%%%%%%%%%%%%%%%%%%%%%%%%%%%%%%%%%%%%%%%%%%
\newpage

%%%%%%%%%%%%%%%%%%%%%%%%%%%%%%%%%%%%%%%%%%%%%%%%%%%%%%%%%%%%%%%%%%%%%%%%%%%%%%%%%%%%%%%%%%%%%%%%%%%%
%%%%%%%%%%%%%%%%%%%%%%%%%%%%%%%%%%%%%%%%%%%%%%%%%%%%%%%%%%%%%%%%%%%%%%%%%%%%%%%%%%%%%%%%%%%%%%%%%%%%

%%%%%%%%%%%%%%%%%%%%%%%%%%%%%%%%%%%%%%%%%%%%%%%%%%%%%%%%%%%%%%%%%%%%%%%%%%%%%%%%%%%%%%%%%%%%%%%%%%%%

%%%%%%%%%%%%%%%%%%%%%%%%%%%%%%%%%%%%%%%%%%%%%%%%%%%%%%%%%%%%%%%%%%%%%%%%%%%%%%%%%%%%%%%%%%%%%%%%%%%%%
%Dead cone effect
% Y.~L.~Dokshitzer{\it~et al.},~Phys. Lett.~B\textbf{519},~199,~(2001).
\end{document}